\documentclass[twocolumn,aps,prd,nofootinbib,floatfix,preprintnumbers,superscriptaddress]{revtex4-1}
\usepackage{graphicx}
\usepackage{color}

\usepackage[normalem]{ulem}

\def\alt{\raise0.3ex\hbox{$\;<$\kern-0.75em\raise-1.1ex\hbox{$\sim\;$}}}
\def\agt{\raise0.3ex\hbox{$\;>$\kern-0.75em\raise-1.1ex\hbox{$\sim\;$}}}

\definecolor{Black}{named}{Black}
\definecolor{Red}{named}{Red}
\definecolor{Green}{named}{Green}

\newcommand{\bw}{\begin{widetext}}
\newcommand{\ew}{\end{widetext}}

\begin{document}

\title{Joint anisotropy and source count constraints on the contribution of blazars to the diffuse gamma-ray background}
\author{A.~Cuoco}
\affiliation{The Oskar Klein Centre for Cosmo Particle Physics, AlbaNova,
SE-106 91 Stockholm, Sweden}
\author{E.~Komatsu}
\affiliation{Texas Cosmology Center and Department of Astronomy, The University of
Texas at Austin, Austin, TX 78712 USA}
\author{J.~M.~Siegal-Gaskins}
\affiliation{Einstein Fellow}
\affiliation{California Institute of Technology, Pasadena, CA 91125 USA}
\affiliation{Center for Cosmology and Astro-Particle Physics, The Ohio State University, Columbus, OH 43210 USA}

\begin{abstract}
We place new constraints on the contribution of blazars to the large-scale isotropic gamma-ray background (IGRB) by jointly analyzing the measured source count distribution ($\log N$-$\log S$) of blazars and the measured intensity and anisotropy of the IGRB\@.  We find that these measurements point to a consistent scenario in which unresolved blazars make $\lesssim 20$\% of the IGRB intensity at 1--10~GeV while accounting for the majority of the measured anisotropy in that energy band.  These results indicate that the remaining fraction of the IGRB intensity is made by a component with a low level of intrinsic anisotropy. We determine upper limits on the anisotropy from non-blazar sources, adopting the best-fit parameters of the measured source count distribution to calculate the unresolved blazar anisotropy.  In addition, we show that the anisotropy measurement excludes some recently proposed models of the unresolved blazar population.
\end{abstract}

\date{\today}

\maketitle

\section{Introduction}
The origin of the isotropic gamma-ray background (IGRB),
the observed all-sky diffuse emission at MeV to GeV energies,
 remains uncertain.    Some or all of this emission is expected to arise from 
astrophysical sources
(e.g., active galactic nuclei, blazars, star-forming galaxies, millisecond pulsars, galaxy clusters, cluster shocks, and cascades from ultra-high-energy cosmic rays; see~\cite{Dermer:2007fg}) as well as possible exotic sources (e.g., dark matter annihilation or decay~\citep{Bergstrom:2001jj}).  The \emph{Fermi} Large Area Telescope (LAT) Collaboration has provided a new 
measurement of the IGRB energy spectrum~\citep{Abdo:2010nz} with
improved accuracy, covering the broad energy range from 200~MeV to
$\sim$~100~GeV.  However, the energy spectrum of the IGRB is reported to be
consistent with a power law, and thus provides little clue to the origin of this emission
in the form of spectral features.  As a result, the contributions of
individual source classes to the IGRB are poorly constrained,
severely limiting our ability to search for signals of new physics or
to place constraints on emission from exotic sources.

One way to tackle the problem is through population studies of resolved sources.  Gamma-ray source classes with resolved members, such as blazars and millisecond pulsars, are obvious candidate contributors to the IGRB via emission from their yet unresolved members. 
The source count distribution in flux ($\log N$-$\log S$,
the number of sources, $N$, per unit flux, $S$) of LAT-detected gamma-ray blazars has recently been studied~\citep{Collaboration:2010gqa} down to fluxes of $S_{100} \sim 10^{-10}$ cm$^{-2}$s$^{-1}$, where $S_{100}$ denotes the individual source flux above 100 MeV\@.  For the first time, the $\log N$-$\log S$ is found to be well-described by a \emph{broken} power law, and the position of the break and the slope of the $\log N$-$\log S$ below and above the break have been measured.  

It is possible to estimate the 
contribution to the IGRB of blazars below the LAT point source
sensitivity of $\sim 10^{-10}$ cm$^{-2}$s$^{-1}$ by extrapolating the
measured $\log N$-$\log S$ to lower fluxes.  
For the energy range 0.1--100 GeV,
Ref.~\citep{Collaboration:2010gqa} reports that unresolved point sources
contribute $22.5\pm1.8$\% of the IGRB intensity measured
by~\citep{Abdo:2010nz}.  As blazars constitute the vast majority of
LAT-detected sources, this is a good indicator of the expected unresolved blazar
contribution, as well as 
a firm upper limit for sources that follow the measured source count distribution.  In the following we use the terms blazars and unresolved sources interchangeably.

\begin{figure*}[t]
\centering
\includegraphics[width=7.8cm]{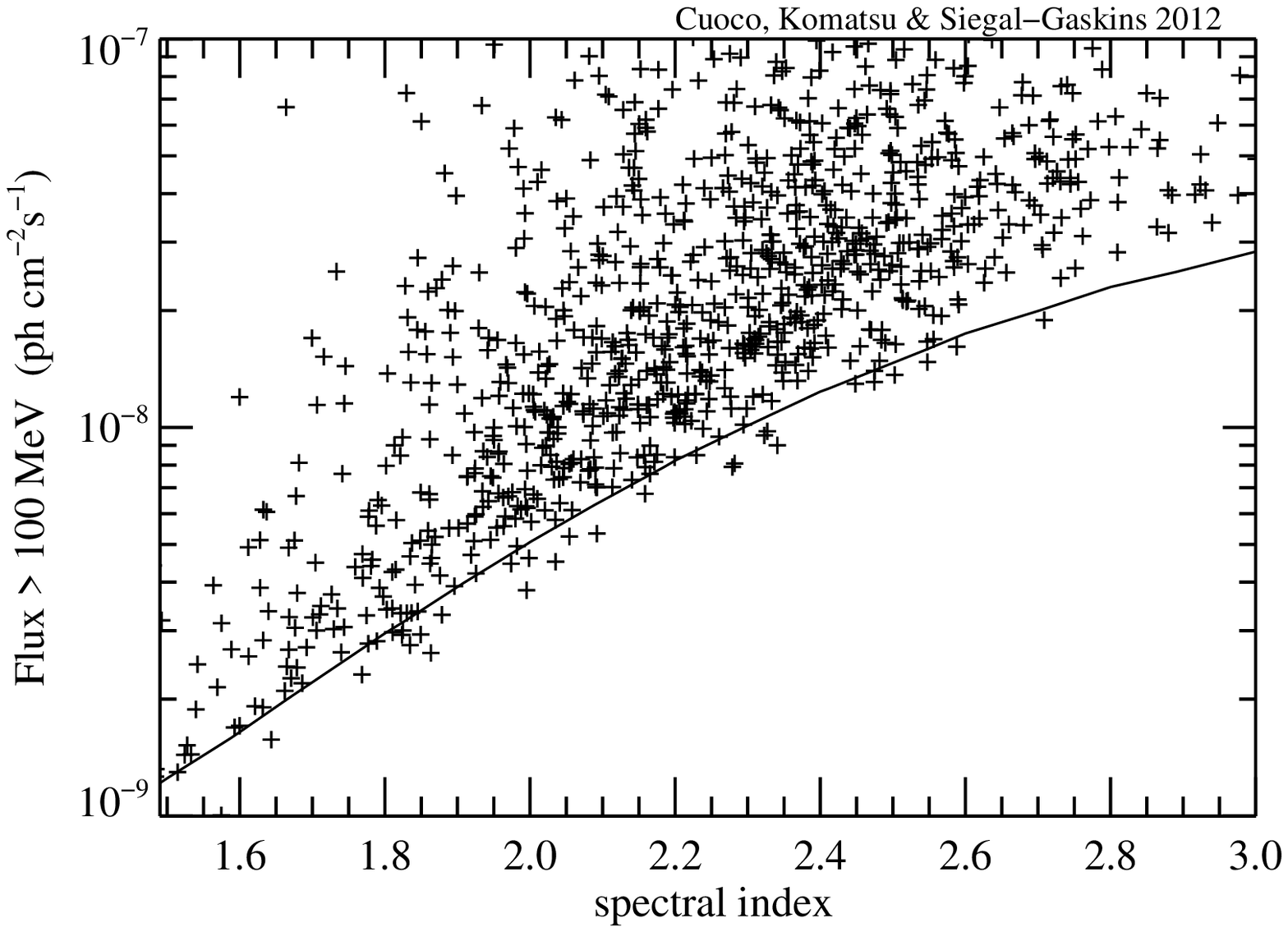}  \hspace{0.5cm}
\includegraphics[width=7.8cm]{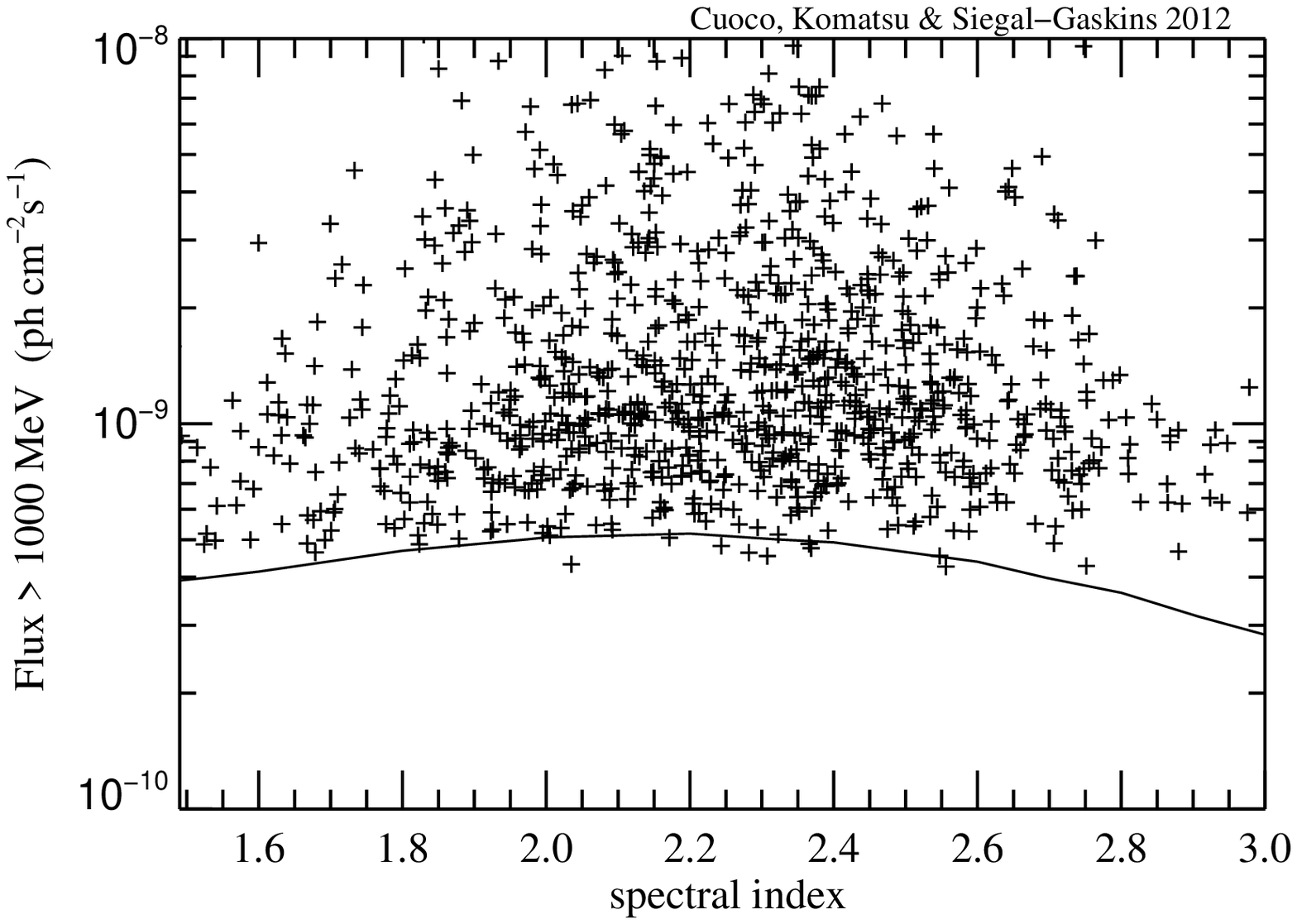} 
\vspace{-0.9pc}
\caption{Left: Spectral index vs.~integrated flux above 100 MeV for the sources in the 1FGL catalog~\cite{1FGL} located at Galactic latitudes $|b| > 10^\circ$. Right: the same for an integrated flux above 1 GeV. The bold lines denote an analytical calculation of the detection threshold (see text for more details).
\label{fig:scatter}}
\end{figure*}

Another constraint, which has not yet been explored, is provided by the level of anisotropy of the IGRB\@.  Recently the first measurement of the small-scale anisotropy of the IGRB has been
made~\citep{anipaper}, while in the last few years predictions have been derived for the anisotropy of many gamma-ray source classes, including blazars and galaxy clusters~\citep{clusters}, millisecond pulsars~\citep{SiegalGaskins:2010mp}, star-forming galaxies~\citep{Ando:2009nk} and dark matter annihilation and decay~\citep{DManisotropies}.
These source classes often produce similar energy spectra but very different anisotropies, suggesting that anisotropy analysis could be a powerful tool for distinguishing possible IGRB contributors.  

In this work, for the first time we use the observed anisotropy information to constrain the properties of the source classes contributing to the IGRB\@.
We calculate the intensity and anisotropy produced by the unresolved
members of a source population whose detected members follow a broken
power-law $\log N$-$\log S$ (such as the LAT-detected blazars). We determine upper limits on the IGRB anisotropy from non-blazar sources by subtracting the predicted angular power for the best-fit blazar $\log N$-$\log S$ from the total measured anisotropy in several energy bands~\citep{anipaper}.  We then allow the $\log N$-$\log S$ parameters to vary, and compare the predictions for this class of models with the measured IGRB intensity~\citep{Abdo:2010nz} and anisotropy~\citep{anipaper} to identify the range of $\log N$-$\log S$ parameters that are consistent with both measurements.

\section{The source count distribution}
\label{sec:srcct}
The $\log N$-$\log S$ of sources detected by the LAT is compatible with a broken power law~\citep{Collaboration:2010gqa},
\begin{equation}\label{LogNLogS}
 \frac{dN}{dS} =  
\left\{
\begin{array}{cc}
   A~{S}^{-\beta} & S\geq S_{\rm b}  \\
   A~{S_{\rm b}} ^{-\beta+\alpha} S^{-\alpha}  &   S<S_{\rm b}
\end{array}
\right.
\end{equation}
where $A$ is the normalization, $S_{\rm b}$ is the flux where the power law breaks, and $\alpha$ and $\beta$ are the power-law slopes below and above the break, respectively. 
The fluxes $S$ and $S_{\rm b}$ are implicitly normalized to  \mbox{1 cm$^{-2}$ s$^{-1}$}.

The $\log N$-$\log S$ of the \emph{Fermi} LAT sources has been measured in several energy bands~\citep{Collaboration:2010gqa}. For consistency with the anisotropy measurements, which were performed only above 1~GeV, we use the best-fit parameters for the $\log N$-$\log S$ in the energy band 1--10~GeV: $A= 
(3.61\pm 0.17) \times 10^{-11}$ cm$^2$ s sr$^{-1}$, $S_{\rm b}= (0.23 \pm 0.06)\times 10^{-8}$ cm$^{-2}$ s$^{-1}$, $\alpha=1.52 \pm 0.15$ and $\beta=2.38 \pm 0.15$~\citep{Collaboration:2010gqa,majello}.  In the following, $S$ denotes the integrated source flux in the 1--10~GeV energy band.
  
From a given $\log N$-$\log S$, the contribution to the
IGRB intensity from the unresolved sources can be estimated by
integrating the distribution from the source detection threshold down to zero flux:
\begin{equation}\label{LogNLogSflux}
I=\int_0^{S_{\rm t}}  \frac{dN}{dS} S~dS,
\end{equation}
where $S_{\rm t}$ is the flux sensitivity threshold for point source detection.  We note that the source detection threshold can in principle depend on a variety of factors.  Of particular relevance for this study, due to the energy-dependent angular resolution of the LAT, the value of $S_{\rm t}$ for a given source depends on its spectral index. However, this \emph{spectral index bias}  is small in the 1--10~GeV range (see next section and~\cite{Collaboration:2010gqa}).
We can thus use an effective $S_{\rm t}$ as described in the next section.

We also note that, concerning the predicted contributions to the IGRB intensity and anisotropy, the extrapolation of the $\log N$-$\log S$ down to zero flux is not strictly necessary for the models we consider.  Since the contribution per logarithmic interval to the IGRB intensity behaves like $S^{(-\alpha+2)}$ and, for $\alpha<2$ (as we consider here), peaks at $S_{\rm b}$ which is close to the detection threshold, the result is insensitive to the exact behavior of the $\log N$-$\log S$ at very low fluxes and to the exact lower flux limit of the extrapolation.
Similar considerations apply to the IGRB anisotropy contribution.
The total number of sources, on the other hand, may be more sensitive to the extrapolation of the $\log N$-$\log S$ at low fluxes. However, this quantity is unobservable unless future instruments are able to resolve all of the IGRB into sources, and, in any case, does not impact the results of the present analysis.

\section{Flux threshold}

Fig.\ref{fig:scatter} shows the spectral index vs.~integrated flux above 100 MeV (left panel) and above 1 GeV (right panel) for the sources in the 1FGL catalog~\cite{1FGL} located at a Galactic latitude of $|b| > 10^\circ$.
The plot clearly shows the strong spectral index bias present in the source fluxes above 100 MeV. On the other hand it is also evident that the spectral index bias is absent or, at most, very weak when considering fluxes above 1 GeV.  Since our analysis focuses on the energy range 1-10 GeV,   we adopt a single threshold independent of the spectral index.  We can derive the flux threshold $S_{\rm t}$ directly from Fig.~\ref{fig:scatter} since the integrated flux of a source in the energy range 1-10 GeV is, to very good approximation, equal to its flux integrated above 1 GeV.
We can see that no sources below a flux of $S=4.0 \times 10^{-10}$ cm$^{-2}$ s$^{-1}$ are detected, while the number of detected sources decreases abruptly below approximately $S=6.0 \times 10^{-10}$ cm$^{-2}$ s$^{-1}$.  In the following we thus adopt the threshold $S_{\rm t}=5.0 \times 10^{-10}$ cm$^{-2}$ s$^{-1}$.

For comparison, we also perform an analytic calculation of the flux threshold as a function of the spectral index following the prescription given in Appendix A of~\cite{1FGL}.  For this calculation we assume a spatially uniform background given by the average of the official Fermi background  model for $|b| > 10^\circ$.  We assume an observation time of 11 months and a detection threshold test statistic TS=25, appropriate for the 1FGL catalog described in \cite{1FGL}. 
Overall, the analytic calculation, shown as bold lines in Fig.~\ref{fig:scatter}, matches well the behavior of the observed source fluxes as a function of the spectral index. Note that we do not implement the correction described in~\cite{1FGL} to account for source confusion, so the analytic calculation slightly underestimates the detection threshold at high spectral indices (very soft sources) where this effect is more important, and indeed this is evident in Fig.~\ref{fig:scatter}.
 
Note that in the following we will compare theoretical predictions of IGRB intensities integrated in the range 1-10 GeV with the experimental measurement given in \cite{Abdo:2010nz} for which sources from a preliminary version of the 1FGL catalog based on 9 months of data were considered, instead of the final 1FGL source list. However, the differences between the two source lists is very small \cite{Markus}, thus we expect that the threshold we adopted is appropriate.
Furthermore, any small differences between the IGRB intensities derived using the two catalogs are likely smaller than the IGRB measurement error itself, which is dominated by systematic uncertainties in the effective area and the level of residual cosmic-ray contamination.

\section{The angular power spectrum}
The Poisson term of the angular power spectrum of the sources, $C_{\rm P}$, can be calculated from the $\log N$-$\log S$. It takes the same value at all multipoles and is given by
\begin{equation}\label{LogNLogSCp}
C_{\rm P} =      \int_{{0}}^{S_{\rm t}}    \frac{dN}{dS} S^2 ~dS.
\end{equation}
This formula gives $C_{\rm P}$ in the units appropriate for the angular
power calculated from an intensity map, i.e., units of intensity$^{2}$
$\times$ solid angle, where intensity is in units
of the number of photons per area per time per solid angle.
Evaluating
Eq.~\ref{LogNLogSCp} using the best-fit $\log N$-$\log S$ parameters and our adopted $S_{\rm t}$ yields $C_{\rm
P,\rm pred} = 1.12 \times 10^{-17}$ (cm$^{-2}$ s$^{-1}$ sr$^{-1}$)$^{2} \, $sr.
Under the assumption that the sources are point-like and unclustered on
the angular scales of interest, the Poisson term is the only
contribution to the angular power.

The predicted power, $C_{\rm P,\rm pred}$,
 has an uncertainty due 
 to the propagation of the uncertainties in the  parameters of the $\log
 N$-$\log S$ function. We estimate this uncertainty from
 the  $\log N$-$\log S$ in the range 0.1--100~GeV, where the parameters
 have smaller statistical uncertainties, and then rescale it to the
 1--10~GeV range. The rescaling was done by
 computing 
the flux conversion factor between the two different bands:
 $\kappa=S_{x\!-\!y}/S_{u\!-\!v}=\left( y^{-\gamma+1}- x^{-\gamma+1}
 \right)/\left( v^{-\gamma+1}- u^{-\gamma+1} \right)$, where $\gamma$ is
 the average photon index of the sources, and $x$$-$$y$ and $u$$-$$v$
 are the edges of the two energy bands.   Then the Poisson anisotropies
 in the two bands are simply related by $C_{\rm P}^{uv}=\kappa^2C_{\rm
 P}^{xy}$.  From the full covariance matrix~\citep{majello} of  $\log
 N$-$\log S$ parameters in the 0.1--100~GeV range, we obtain $\delta C_{\rm P}^{0.1\!-\!100}= 0.65 \times 10^{-15}$ (cm$^{-2}$ s$^{-1}$ sr$^{-1}$)$^{2}$ sr, which, assuming $
 \gamma=2.4$, can be rescaled to  $\delta$$C_{\rm P}^{1\!-\!10} \equiv \delta C_{\rm P,pred} = 0.95 \times 10^{-18}$ (cm$^{-2}$ s$^{-1}$ sr$^{-1}$)$^{2}$ sr, so that  $C_{\rm P,pred} = (11.2\pm1.0) \times 10^{-18}$ (cm$^{-2}$ s$^{-1}$ sr$^{-1}$)$^{2}$ sr.

\begin{figure}[t]
\includegraphics[width=8.5cm]{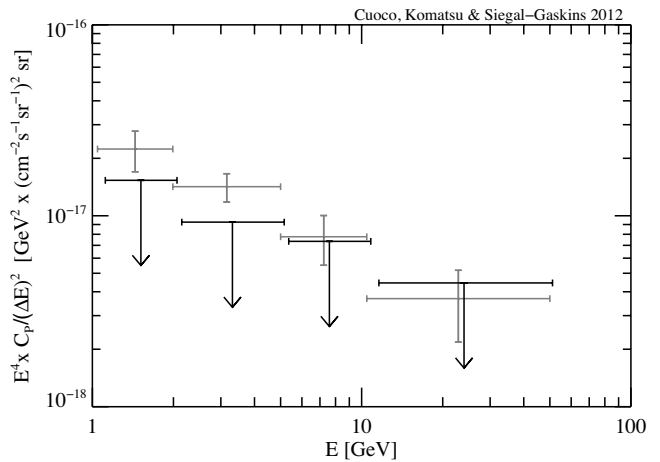}  
\vspace{-0.9pc}
\caption{Anisotropy measurements with 1$\sigma$ uncertainties (\emph{light gray crosses}) from Ref.~\citep{anipaper} and 2$\sigma$ upper limits (\emph{black bars}) from this work on the anisotropy from non-blazar components in different energy bands.  
The points  have been divided by the square of the energy bin width $(\Delta E)_i^2$ and multiplied   by $E_i^4$, with $E_i$ the logarithmic center of the energy bin. 
The upper limits are slightly displaced in energy for clarity.\label{fig:ulimits}}
\end{figure}

\begin{table}[b]
\vspace{-0.5pc}
\caption{\label{tab:ulims}
Measured angular power
from the foreground-cleaned data in
 different energy bands, $C_{\rm P, data}$~\citep{anipaper}; predicted
power from 
 unresolved blazars, $C_{\rm P,
 pred}$  (this work); and 2$\sigma$ upper limits on the residual
 anisotropy, $C_{\rm P, U}^{2\sigma}$ (this work).}  
\begin{ruledtabular}
\begin{tabular}{ccccc}
\multicolumn{1}{c}{$E_{\rm min}$}&
\multicolumn{1}{c}{$E_{\rm max}$}&
\multicolumn{1}{c}{$C_{\rm P, data}$}&
\multicolumn{1}{c}{$C_{\rm P, pred}$}&
\multicolumn{1}{c}{$C_{\rm P, U}^{2\sigma}$} \\
\multicolumn{1}{c}{\textrm{[GeV]}}&
\multicolumn{1}{c}{\textrm{[GeV]}}&
\multicolumn{3}{c}{[$10^{-19}$ (cm$^{-2}$ s$^{-1}$ sr$^{-1}$)$^{2}$ sr]}
\\
\colrule
  	1.0  &	 10.0 & 	$110 \pm 12 $ & 		$112 \pm 10 $ &  	            $<33   $   	\\
\colrule
 	1.04 & 	1.99 & 	$46.2 \pm 11.1$ & 		$38.0 \pm 3.9 $   &	  $<32 $   	\\
 	1.99 & 	5.00 & 	$11.30 \pm 2.20$ & 		$9.3 \pm 0.9 $     &   	  $<6.7  $   	\\
  	5.00 & 	10.4 & 	$0.845 \pm 0.246$ &   	$0.55 \pm 0.05 $ &	  $<0.80  $   	\\
  	10.4 &	 50.0 & 	$0.211 \pm 0.086 $ & 	$0.13 \pm 0.01 $ &      $<0.254 $   	\\
\end{tabular}
\end{ruledtabular}
\end{table}

\begin{figure*}[t]
\centering
\includegraphics[width=8.2cm]{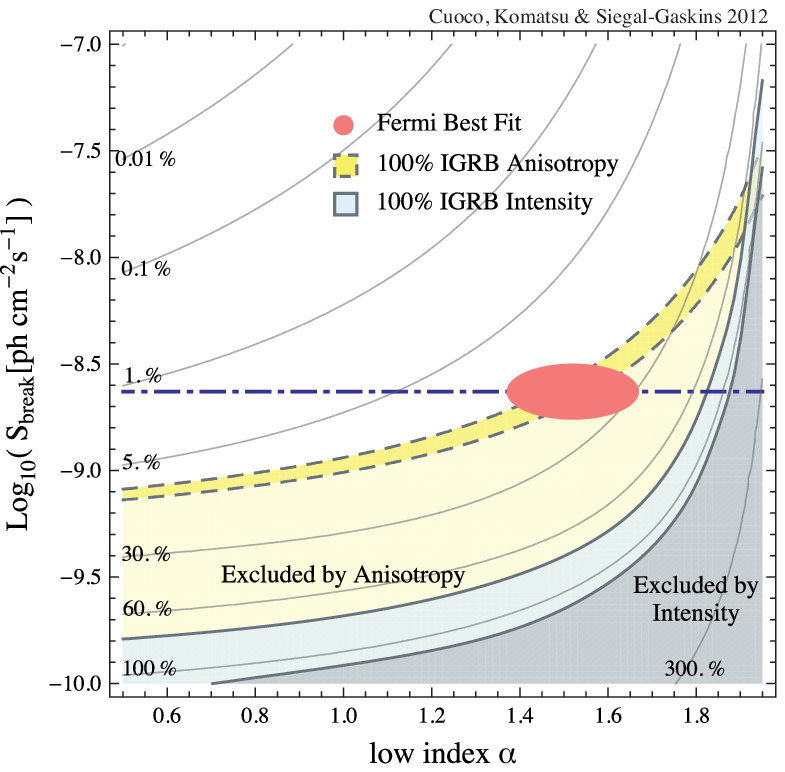}  \hspace{0.5cm}
\includegraphics[width=8.2cm]{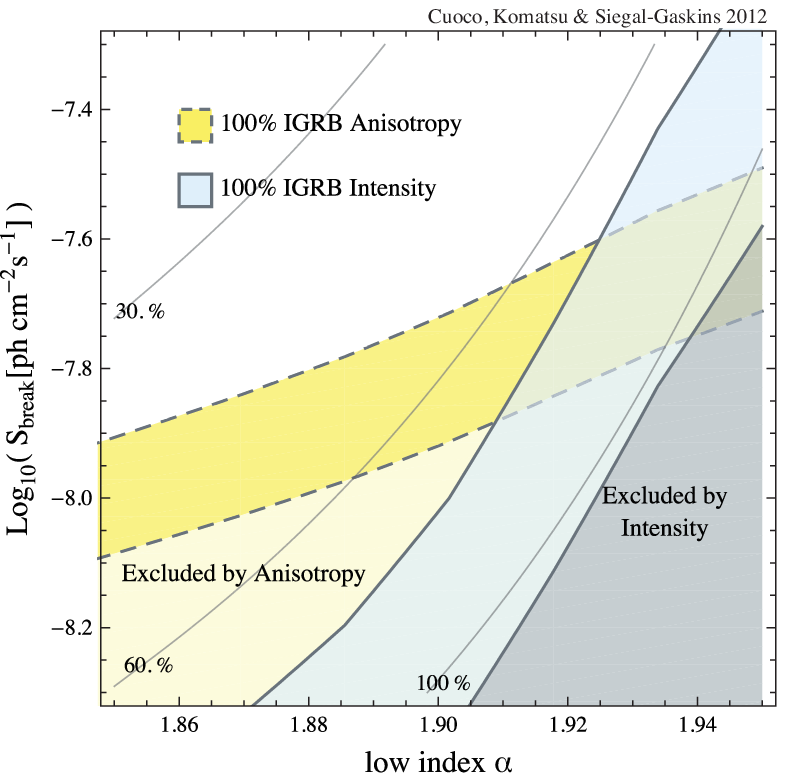}  
\vspace{-0.9pc}
\caption{Left: Constraints on blazar $\log N$-$\log S$
 parameters (break flux, $S_{\rm b}$, and faint-end slope,
 $\alpha$) from the intensity and anisotropy of the IGRB\@.  
Regions in which blazars provide 100\% of the observed IGRB
 anisotropy and mean intensity in the 1--10~GeV energy
 band are shown; the widths of the regions indicate the 68\% confidence
 intervals. 
 Below these regions blazars overproduce the anisotropy and mean
 intensity.  
 Labeled contours show 
the fraction of the blazar contribution to the IGRB intensity.
The best-fit 1$\sigma$ parameter
region from the \emph{Fermi} source count analysis~\citep{Collaboration:2010gqa}  
is marked, along with the best-fit $S_{\rm b}$~\citep{Collaboration:2010gqa} (dot-dashed line).  
Right: expanded view around the region of parameter space in the left panel where blazars contribute 100\% of both the measured IGRB anisotropy and intensity.
\label{fig:contours}}
\end{figure*}

Ref.~\citep{anipaper} reports measurements of $C_{\rm P}$, obtained by averaging the angular power spectrum coefficients $C_{\ell}$ over the multipole range $155 \le \ell \le 504$, in the energy ranges 1--2~GeV, 2--5~GeV, 5--10~GeV, and 10--50~GeV.   Using the same analysis pipeline as Ref.~\cite{anipaper}, we have also calculated the anisotropy for the 1--10~GeV energy band for the \emph{foreground-cleaned} data, which yields  $C_{\rm P,data} = (11.0 \pm 1.2) \times 10^{-18}$ (cm$^{-2}$ s$^{-1}$ sr$^{-1}$)$^{2}$ sr.  This value can be directly compared with the predicted value derived above. The two values are compatible at the 1$\sigma$ level so that unresolved blazars can account for all of the observed anisotropy.  We discuss this point further in the next section.  
The 2$\sigma$ upper limit on the non-blazar anisotropy
is $C_{\rm P, U}^{2\sigma} = (C_{\rm P, data}-C_{\rm P, pred})+2\times \sqrt{\delta C_{\rm P, data}^{2}+\delta C_{\rm P, pred}^{2}} = 3.3 \times 10^{-18}$ (cm$^{-2}$ s$^{-1}$ sr$^{-1}$)$^{2}$ sr.

Using the best-fit $\log N$-$\log S$, we also compare the
predicted $C_{\rm P}$ with the anisotropy measurements in the four
energy bands used in Ref.~\citep{anipaper}  
(Table~\ref{tab:ulims}). 
In this case we use the rescaling method described above to calculate both  the predicted mean values and their uncertainties.
The derived 2$\sigma$ upper limits on the level of residual anisotropy in each energy bin are  reported in 
Table~\ref{tab:ulims} and shown in Fig.~\ref{fig:ulimits}.
These limits can be used to constrain models of astrophysical
or exotic source populations, based on their predicted
level of anisotropy.  
We note that the uncertainties, and, except for the 1--10 GeV case, the central values for $C_{\rm P,pred}$ used to derive these limits
rely on the rescaling method described above, and thus on the assumption
of an average index for the sources. However, we find that varying
$\gamma$ from 2.2 to 2.6 produces only a
small change of order $\sim10\%$.

Finally, as a technical remark, we emphasize  that the use of the \emph{dimensionful} intensity angular power, rather than the \emph{dimensionless} fluctuation angular power, conveniently avoids the need to treat contamination of the anisotropy measurement by possible residual Galactic diffuse emission or instrumental backgrounds.  
These backgrounds are, to good approximation, isotropic, or vary only
on large angular scales, and thus their contribution to the intensity
angular power spectrum appears only at multipoles far below the range
used to measure the angular power reported in~\citep{anipaper}.  As stated previously, in the
following, when discussing the IGRB intensity $I_{\rm IGRB}$ we
consider the measurement given in~\citep{Abdo:2010nz}.

\section{Constraints on unresolved blazars}
We now 
explore more generally the parameter space of the $\log N$-$\log
S$ function to determine the region that is compatible with
the measured anisotropy, intensity, and source count
data.  We define the parameter space of the source count
distribution by the position of the break flux, $S_{\rm b}$, and the
faint-end slope, $\alpha$, of the $\log N$-$\log S$ function at
fluxes below the break flux.
We fix the normalization and slope of the $\log N$-$\log S$ at high
fluxes, as the efficiency in detecting point sources at
high fluxes is $\sim 1$, and thus these parameters are well-determined
(i.e., potential biases in these parameters are small).
For each point in the $S_{\rm b}$-$\alpha$ parameter space we calculate
the predicted $I_{\rm IGRB}$ and $C_{\rm P}$ from the corresponding
$\log N$-$\log S$ function.  

In Fig.~\ref{fig:contours} we show the region of the $\log
N$-$\log S$ parameter space in which blazars contribute 100\% of the IGRB intensity (light blue) and that
in which they contribute 100\% of the angular power (dark yellow) in the 1--10~GeV energy band.  The widths
of these regions show the {68}\%~(1$\sigma$) confidence level regions,
reflecting the respective 1$\sigma$ uncertainties in the measured
$C_{\rm P}$ and $I_{\rm IGRB}$.  
 Above the light blue region,
 blazars contribute less than 100\% of the measured IGRB intensity;
 below this region,
 blazars overproduce the IGRB intensity.  
Similarly, 
above the dark yellow
region blazars do not contribute the entirety of the
measured angular power, whereas below this region they overproduce the anisotropy.
We emphasize that the constraint  
from the anisotropy measurement is much stronger than that from the intensity
measurement except for at very high values of $\alpha$.
 
There is a region of parameter space in which blazars contribute 100\%
of the IGRB intensity without exceeding the measured
$C_{\rm P}$; however, this region (shown expanded in the right panel of Fig.~\ref{fig:contours})  has   
a high break flux ($S_{\rm b} \approx$~few~times~$10^{-8}$cm$^{-2}$ s$^{-1}$) which
is strongly incompatible with the break measured from the source
count analysis. 
Such a high break flux can be robustly excluded, as it
would lie in the flux range where the source detection
efficiency is close to 1,  
and thus this kind of feature is unlikely to have been missed.
Taking
the measured value of the break flux 
as an upper limit, we find that the contribution from blazars in the region allowed by
$C_{\rm P,data}$  
cannot be more than $\sim$20\% of the IGRB mean intensity (see labeled contours in Fig.~\ref{fig:contours}),
a value which is in 
agreement with the results of the source count analysis alone.

We now ask
``how well do the parameters of the $\log N$-$\log S$ function inferred
from $C_{\rm P, data}$ agree with those found from the source
count analysis?'' 
The best-fit
1$\sigma$ region 
of $S_{\rm b}$ and $\alpha$ for the blazar
$\log N$-$\log 
S$ given in \cite{Collaboration:2010gqa,majello} overlaps well with the 1$\sigma$ region where blazars contribute 100\% of
$C_{\rm P,data}$.
This is a non-trivial result, as the measured anisotropy and source count distribution are independent observables, determined from independent data analyses.  
Note that the errors on the 1--10 GeV $\log N$-$\log S$ parameters shown in the plot are  taken directly from \citep{Collaboration:2010gqa,majello} and are  larger than the rescaled ones used in the previous section.

To further demonstrate how anisotropy data can be a powerful tool for distinguishing
between multiple scenarios 
we test an alternative fit to the
blazar $\log N$-$\log S$ obtained by Stecker \& Venters~\cite{Stecker:2010di}. A notable
feature of this alternative fit is that it can account for $\sim$60\% of the
IGRB intensity in the 1-10~GeV energy band. 
We  have  calculated $C_{\rm P}$ from the $\log N$-$\log S$ of the Stecker \& Venters model~\cite{Stecker:2010di,venters} and, using 
a threshold of $5.0 \times 10^{-10}$ cm$^{-2}$ s$^{-1}$ (the same used in the rest of our analysis),
obtain $C_{\rm P}=(3.0\pm0.5) \times 10^{-17}$ (cm$^{-2}$ s$^{-1}$ sr$^{-1}$)$^{2}$ sr (the error reported on this prediction being likely an overestimate since it neglects the covariance of the parameters).  This value is a factor of $\sim 3.0$ larger than the measured value, and is inconsistent with $C_{\rm P,data}$ at $3.7\sigma$.  The anisotropy data thus strongly excludes this blazar model.  In addition, we remark that the recent analysis of \cite{Harding:2012gk} using the blazar model of \cite{Abazajian:2010pc} reaches conclusions similar to those of the present study: those authors find that the measured IGRB anisotropy places a strong constraint on the contribution of blazars to the intensity of the IGRB, and that, assuming the model considered in that work, blazars cannot contribute a substantial fraction of the IGRB intensity.
 
Comparing the measured anisotropy of the IGRB and the predicted anisotropy from blazars leads to another important conclusion.
Since, for the best-fit source count distribution, blazars already account for $\sim 100$\% of the observed anisotropy and, in intensity units, Poisson angular power is additive, the remaining component (or components) making $\sim 80$\% of the IGRB intensity must contribute a  low level of anisotropy in order to not overproduce the observed angular power.  Interestingly, this can be achieved quite naturally since some proposed contributors to the IGRB, such as star-forming galaxies~\cite{Ando:2009nk}, are expected to contribute negligibly to the anisotropy.  On the other hand, this result implies strong constraints on source populations with large intrinsic anisotropy.

\begin{figure}[t]
\centering
\includegraphics[width=8.2cm]{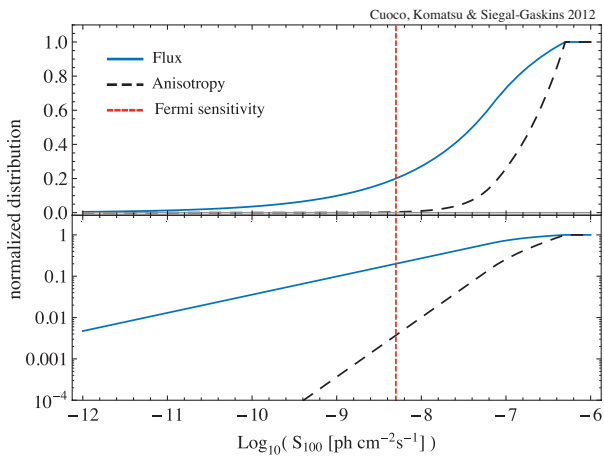}  
\vspace{-0.7pc}
\caption{Cumulative contribution of blazars in linear (top) and log (bottom) scale to
 the IGRB anisotropy (dashed) and intensity (solid) for the \emph{Fermi} best-fit $\log N$-$\log S$ ($E>100$~ MeV) as a function of source flux.
\label{fig:differential}}
\end{figure}

We emphasize that the anisotropy and intensity contributions from a source population have different dependences on the source count distribution, and consequently they represent complementary observables which are sensitive to different source flux ranges.
This is demonstrated in Fig.~\ref{fig:differential}, which shows the cumulative contribution to the intensity and anisotropy  above 100~MeV as a function of source flux for the \emph{Fermi} LAT best-fit $\log N$-$\log S$ parameters.  From the relative flatness of the cumulative flux distribution
below the threshold flux, it can be inferred that the IGRB intensity contribution from unresolved blazars has only a weak dependence on the effective flux sensitivity.  The cumulative anisotropy distribution, however, falls off more quickly below the threshold flux, so the anisotropy from unresolved sources is more strongly dependent on the sensitivity limit, and improved point source sensitivity is thus likely to have a more notable impact on the measured IGRB anisotropy.  

\section{Additional energy bands} \label{sec:morebands}
We briefly consider this analysis in other energy bands.  
The range above 10~GeV is currently not suitable since the error on the measured $C_{\rm P}$ is large and the $\log N$-$\log S$ is not well-constrained. 
A natural extension is thus to include the low-energy range down to 100~MeV.  However, spectral index bias is non-negligible in this energy band, 
and needs to be taken into account. Equations~\ref{LogNLogSflux} and~\ref{LogNLogSCp} in this case also need to be modified as:
\begin{eqnarray}
  I &=&  \int_{\Gamma_{\rm min}}^{\Gamma_{\rm max}}   \!\!\!\!  \int_0^{S_{\rm t}(\Gamma)}  \frac{d^2N}{dSd\Gamma} S~d\Gamma dS \\
  C_{\rm P} &=&  \int_{\Gamma_{\rm min}}^{\Gamma_{\rm max}}    \!\!\!\!   \int_{{0}}^{S_{\rm t}(\Gamma)}     \frac{d^2N}{dSd\Gamma} S^2 ~d\Gamma dS,
\end{eqnarray}
where $d^2N/dSd\Gamma$ is now the source counts per unit flux and unit spectral index, and  $S_{\rm t}(\Gamma)$ is detection threshold as a function of spectral index, which can be calculated as described in \S\ref{sec:srcct}.
However, since there is not yet a measurement of the anisotropy below 1~GeV to confront with the theoretical prediction, we leave a more detailed analysis of lower (and higher) energy bands to future work.
Clearly, a more detailed model of the sources is now required to predict the intensity and anisotropy of the IGRB. The distribution in spectral indices now becomes important, while the calculation was previously insensitive to this property of the sources.
Also, while a simple broken power law was accurate enough to describe $dN/dS$, more parameters are now required to describe the full $d^2N/dSd\Gamma$.  The formalism presented above can be applied to physically-motivated models of IGRB contributors from population synthesis of source classes (as in \cite{Stecker:2010di,Abazajian:2010pc,Inoue:2008pk,Fields:2010bw,Loeb:2000na}), where effective $dN/dS$ and $d^2N/dSd\Gamma$ in different energy bands is a prediction of the model itself, without the need to assume a parametric functional form.  The $C_{\rm P}$ prediction for the Stecker \& Venters blazar model \cite{Stecker:2010di} derived in this work is an example of such an application.

\section{Conclusions}%
We performed a joint analysis of the source count distribution of blazars and the measured anisotropy of the IGRB in the energy range 1--10~GeV, and find that a consistent picture emerges in which unresolved blazars account for only $\sim 20$\% of the IGRB intensity but $\sim 100$\% of the angular power.  The sources contributing the remaining $\sim 80$\% of the IGRB intensity are thus constrained to provide only a small contribution to the anisotropy.  Viable models of sources contributing to the IGRB must satisfy the upper limits on their anisotropy that we reported in the last column of Table~\ref{tab:ulims}.  We recommend that proposed models of sources contributing to the IGRB should be provided in terms of effective $dN/dS$ (and $d^2N/dSd\Gamma$ when appropriate) in different energy bands in order to ease  the use and application of these models by the scientific community, in particular for comparing with the measured IGRB anisotropy, which represents a newly available observable.
These results demonstrate  the power of anisotropy information for constraining the origin of the IGRB\@.

\acknowledgements
We thank M.~Ajello, T.~Venters, F.~Stecker, and K.~Abazajian for useful comments and interesting discussions.  EK is supported in part by NSF grants AST-0807649 and PHY-0758153 and NASA grant NNX08AL43G.  JSG acknowledges support from NASA through Einstein Postdoctoral Fellowship grant PF1-120089 awarded by the Chandra X-ray Center, which is operated by the Smithsonian Astrophysical Observatory for NASA under contract NAS8-03060.


\end{document}